\documentclass{jfm}
\usepackage{graphicx}
\usepackage{epstopdf, epsfig}
\usepackage{amsmath}
\usepackage{cleveref}

\newcommand{\deff}{$D_{\text{eff}}$}

\newcommand{\erf}{\text{Erf}}

\usepackage{color}
 
\definecolor{BluePastel}{RGB}{119,207,255}
\definecolor{RedPastel}{RGB}{255,125,82}

\shorttitle{Solidification dynamic of an impacted drop}
\shortauthor{V. Thi\'evenaz, T. S\'eon and C. Josserand}

\title{Solidification dynamic of an impacted drop}

\author{V. Thi\'evenaz\aff{1},
  T. S\'eon\aff{1}
    \corresp{\email{thomas.seon@gmail.com}}
 \and C. Josserand\aff{2}}

\affiliation{\aff{1}Institut $\partial$'Alembert, UMR 7190, CNRS \& Sorbonne Universit\'es, Paris, France
\aff{2}Laboratoire d'Hydrodynamique (LadHyX), UMR7646 CNRS-Ecole Polytechnique, 91128 Palaiseau CEDEX, France}

\begin{document}

\maketitle

\begin{abstract}
This paper is dedicated to the solidification of a water drop impacting a cold solid surface. 
In a first part, we establish a 1D solidification model, derived from the Stefan problem, that aims at predicting the freezing dynamic of a liquid on a cold substrate, taking into account the thermal properties of this substrate. 
This model is then experimentally validated through a 1D solidification setup, using different liquids and substrates. 
In a second part, we show that during the actual drop spreading, a thin layer of ice develops between the water and the substrate, and pins the contact line at its edge when the drop reaches its maximal diameter.
The liquid film then remains still on its ice and keeps freezing. This configuration lasts until the contact line eventually depins and the liquid film retracts on the ice. 
We measure and interpret this crucial time of freezing during which the main ice layer is built.
Finally, we compare our 1D model prediction to the thickness of this ice pancake and we find a very good agreement. 
This allows us to provide a general expression for the frozen drop main thickness, using the drop impact and liquid parameters.









\end{abstract}


\section{Introduction}\label{sec:intro}



When a liquid drop is put in contact with a cold substrate, either by impact or deposition, the freezing of the liquid can lead to unexpected final shapes. Understanding the coupling between drop impact hydrodynamics and solidification, that builds the frozen structure, is crucial in many different contexts : airplane icing~\citep{Baumert2018}, ice accretion on wires or roadways due to freezing rain \citep{Jones1998}, 3D printing~\citep{Lipson2013}, surface metal coating technology~\citep{Pasandideh-Fard2002a,Fauchais2004}, etc. 
Indeed, although most of the airplane icing configurations concern impact of ice crystal~\citep{Vidaurre09,Roisman15}, the impact of water droplets on subfreezing substrates can be of great interest in the dynamics of icing formation~\citep{Schremb2018}. Today, the optimal design and coating of the surfaces to avoid icing remains an important open problem~\citep{Cao2009, Kreder2016}. 
Freezing rain may cause hazardous conditions for pedestrians and cars or break tree limbs and power lines, and thus may cause immeasurable economic losses \citep{Jones1996}.
In thermal spray deposition, the thickness and geometry of the solidified splat is important for the quality of the coating and depends on the melting of the spray with the substrate~\citep{Chandra09}. Similarly, multiple droplet impacts, coupling the fluid dynamics with the solidification thermal processes, produce complex splat patterns that determines the coating quality~\citep{Chandra05}. 

In that context, it is important to have a precise characterization of the thickness of the residual solid layer. It could for instance help to improve the existing ice accretion models, crucial for aircraft icing or ice load on transmission lines \citep{Schremb2018}, or increase coating efficiency formed by the impact and solidification of molten thermal spray particles \citep{Dhiman2007}.
Moreover, because of thermal contraction when the solid layer becomes cooler, and depending on the thickness of the frozen impacted drop and the substrate temperature, the frozen structure can either remain stuck on the substrate or detach through a self-peeling process~\citep{Ruiter2018} or even fragment into a myriad of small ice pieces~\citep{Ghabache2016b}.

When the drop is simply deposited on the substrate, the frozen drop shape and thickness depend on the contact line solidification dynamics~\citep{De-Ruiter2017, Tavakoli2014, Schiaffino1997a}. 
In this paper, we aim at obtaining a prediction of the final ice layer thickness resulting from the impact and solidification of a drop on cold surfaces.
This imposes to consider the complex coupling between freezing and hydrodynamics. Indeed, right after impact, while the drop spreads on the substrate \citep{Josserand2016}, a thin solid crust layer forms between the liquid and the substrate~\citep{Ruiter2018}. Afterwards, as the solid layer keeps growing from the substrate \citep{Marin2014, Gao1994}, the remaining liquid can retract on its solid layer \citep{Bartolo2005}, preventing the final solid layer to reach a uniform thickness.
For the sake of clarity, this paper will be therefore divided into two parts: in the first part, the solidification dynamics of a liquid suddenly put in contact with a cold substrate will be tackled from a general point of view and, in the second part, the results obtained will be applied to the particular case of a water drop impacting such a cold substrate. 

The paper is organized as follows: 
in the first part, after introducing the problem of a melt freezing on a substrate (section \ref{PresentationPb}), we develop and interpret the associated theoretical model (section \ref{model}), and we compare it to a dedicated model experiment (section~\ref{ExperimentalComparison}).
In the second part, the drop impact experimental setup and the measurement techniques involved are first described (section \ref{sec:exp}), then, we depict and interpret the hydrodynamic behaviour of the water layer pinned at the edge of its growing ice (section \ref{sec:res}), and finally, we present our experimental results relating to the formation of the residual ice layer, in the light of the thermal and hydrodynamic behaviours presented before (section \ref{sec:results}).


\section{Solidification dynamics of a liquid on a substrate}

In this section, we introduce a simplified 1-D solidification model that aims at describing the temperature distribution and the solidification front dynamics in the problem of a liquid film on a cold solid substrate.

\subsection{Presentation of the problem}
\label{PresentationPb}

\begin{figure}
    \includegraphics[width=\hsize]{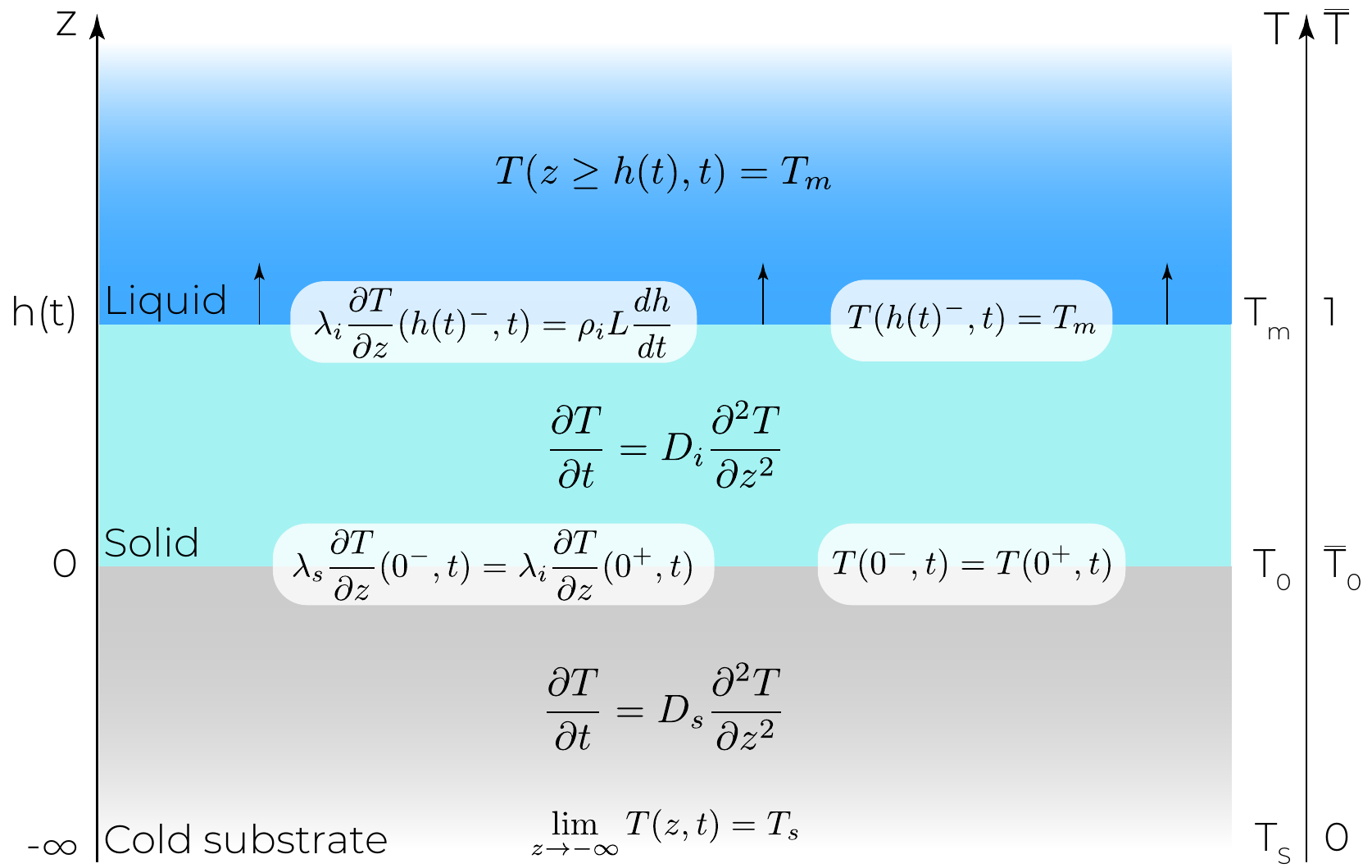}
    \caption{Summary of the model hypotheses:  A finite layer of solid lies between the semi-infinite melt ($z > h(t)$) and the semi-infinite substrate ($z < 0$). The temperature of the whole melt is set constant at the melting point ($T = T_m$), while the temperature of the substrate tends to $T_S$ when $z$ goes to $-\infty$. The temperature in the solid phases is given by a set of two heat equations, with a specific diffusion coefficient $D_k$ for each phase, coupled by the temperature and heat flux continuity at $z=0$. At the solidification front ($z = h(t)$), the Stefan condition imposes the downward thermal flux be equal to the latent heat liberated by the freezing.}
  \label{ModelGeom}
\end{figure}

We consider a material existing in two phase states : liquid and solid.  Then, an infinite flat solid substrate, filling the halfspace $z < 0$, with a temperature T$_s$ below the melting temperature T$_m$, is in contact with the solid layer of the material (for $0 < z < h(t)$), that is growing in its melt ($z > h(t)$). $h(t)$ is the position of the solidification front (see schematic Fig.~\ref{ModelGeom}).
This configuration belongs to the large class of Stefan problems, a particular kind of boundary value problem where a phase boundary can move with time. It is named after Josef Stefan who first, in 1889, solved the original configuration in which a solidification front propagates between two phases (liquid and solid) of the same material, without substrate \citep{Stefan, Brillouin1930}. In figure~\ref{ModelGeom}, this \textit{original Stefan problem} amounts to replace the substrate by the solid phase. 

Configurations belonging to Stefan problems, even restricted to the solidification-melting phase change, are abundant, in an impressively wide range of application. We can cite as examples, the solidification of the earth, supposed to be molten at the origin, which triggered the first work of interest in this area, by Lam\'e and Clapeyron (1831) \citep{Lame1831}. 
The formation of ice crystals, like snowflakes, that grow out from seeds in an environment of supersaturated water vapour. In this problem the  ingredients of the original Stefan problem are not sufficient, Gibbs-Thomson equation and 3D effects need to be added \citep{Langer1980}. 
The modelling of the dynamics of sea ice on the surface of the polar oceans \citep{Worster2000} or the shape of an icicle \citep{Neufeld2010}, where the liquid motions and the associated convective heat transfert have to be considered. 
\textit{Idem} for various lava flows that cool and gradually solidify until they come to rest \citep{Griffiths2000}. Cooling can occur from the surrounding atmosphere (or water) or from the underlying solid \citep{huppert1989}, and in ancient time, some of the lavas were hotter than today and  were even capable of melting the underlying rock and shaping their own thermal erosion bed \citep{Huppert1986}. 
In industry, Stefan problems are a lot studied in the context of melting or solidification of metals or metal alloys \citep{Viskanta1988}, where the properties of the solid (in particular, its mechanical and thermal properties) are functions of the kinetics of solidification, such that Czochralski crystal growth, used for example in the fabrication of semi-conductor wafers \citep{Nishinaga2014}, laser welding \citep{Cline1977, Allmen2013}, or synthesis of nanoparticles from metal films melting \citep{Font2017}.
Finally, this class of problems is also a fantastic playground for mathematicians \citep{Rubinstein1971, Gupta03}. 
But to the best of our knowledge, we are not aware of a system of equations modeling solid growth by a sudden contact between a liquid and a cold solid substrate.



In the following, we assume the melt stays at rest, at constant temperature everywhere ($T_m$). We neglect the variation of heat capacity and of thermal conductivity with temperature. We also neglect the thermal expansion, and more generally the variation of density. In other words, the thermal parameters of the media are the following: the latent heat of solidification L, the heat capacity $C_{pk}$, the thermal conductivity $\lambda_k$, the heat diffusion coefficient $D_k=\lambda_k/(\rho_kC_{pk})$ and therefore the density $\rho_k$, are taken as constant. The subscript $k$ stands for $l$ in the liquid phase, $i$ in the solid phase and $s$ for the substrate (see Fig.~\ref{ModelGeom}). 

The choice of a one-dimensional model will be discussed in \S \ref{ExperimentalComparison} and is supported by the strong aspect ratio of the impacted drop that will be considered later.
We assume that the liquid is at the melting temperature $T_m$: this approximation will also be discussed in \S \ref{ExperimentalComparison}. It can be justified \textit{a priori} by both, the small thickness of the liquid layer and the little energy needed to cool the water down to its melting temperature : $C_{pl} (T_d-T_m) \sim 4000\times 20 \sim 8 \cdot 10^4$ J$\cdot$kg$^{-1}$, where $T_d$ is the initial temperature of the liquid drop, compared to the latent heat for solidification ($L\sim3\cdot 10^5$ J$\cdot$kg$^{-1}$).

Under these assumptions, the mathematical problem that need to be solved is the following (see Fig.~\ref{ModelGeom}): a constant temperature $T_m$ in the liquid, two heat equations for the temperature field $T(z,t)$ :
\begin{equation}
    \frac{\partial T}{\partial t} = D_s \frac{\partial^2 T}{\partial z^2} \,\,\,\,\,\, {\rm for} \,\, z \leq 0 \,;  \hspace{3 cm}
    \frac{\partial T}{\partial t} = D_{i} \frac{\partial^2 T}{\partial z^2} \,\,\,\,\,\, {\rm for} \,\, 0 \leq z \leq h(t),
    \label{eq:diff}
\end{equation}
and four boundary conditions at the two interfaces :
\begin{equation}
T(0^-,t)=T(0^+,t) \,;  \hspace{4 cm} \lambda_s \frac{\partial T}{\partial z}(0^-,t)=\lambda_i \frac{\partial T}{\partial z}(0^+,t),
 \label{BC:z0}
\end{equation}
which impose both the continuity of the temperature and of the heat flux at the substrate/solid interface ($z=0$), and
\begin{equation}
T(h(t)^-,t)=T_m \,;  \hspace{4 cm} \lambda_i \frac{\partial T}{\partial z}(h(t)^-,t)=\rho_i L \frac{dh}{dt},
 \label{BC:zh}
\end{equation}
which impose both the continuity of the temperature, and the law of motion of the solidification front ($z=h(t)$). This energy conservation law under liquid-solid phase change is called the Stefan condition. Here, it imposes that the solidification front velocity is proportional to the rate at which latent heat can be transported in the solid phase. The solidification front is thus controlled by the diffusion in the solid and the substrate through the Stefan condition.
Finally, we impose a constant temperature $T_s$ far in the substrate : 
\begin{equation}
\lim_{z \rightarrow -\infty} T(z,t)=T_s,
\end{equation}
and we complement this set of equations by the
initial conditions taken at $t=0$:
\begin{equation}
T(z,0)=T_s  \,\,\, {\rm for} \,\,\, z \leq 0 \,\, \,\,{\rm and} \,\,\,\, h(0)=0,
 \label{iniBC}
\end{equation}
indicating that at $t=0$ the liquid is suddenly put in contact with the substrate.

\subsection{Solution of our unidimensional solidification model}
\label{model}

Similarity analysis shows that this diffusive problem exhibits a self-similar solution, with the usual self-similar variable involved in diffusion problems $\eta = \frac{z}{\sqrt{t}}$, even in the presence of the moving solidification front.
In this case, the solidification follows the usual diffusive front growth law: 
\begin{equation}
    h(t) =  \sqrt{D_{\text{eff}}t}
    \label{eq:hsim}
\end{equation}
where $D_\text{eff}$ is the effective diffusion coefficient that determines the growth of the solid layer. 
It is really the quantity of interest that we need to compute and thus relate to the different 
thermal properties of our problem.

Introducing the self-similar variable in the set of \Crefrange{eq:diff}{iniBC}, we obtain the following solutions for the temperature field:
\begin{equation}
  T(z,t) =  T_0+ (T_0-T_s)  \cdot \erf \left(\frac{z}{2\sqrt{D_st}}\right) \,\,{\rm for}\,\, z \leq 0 \,\, {\rm and}  
    \label{eq:tempprofs}
\end{equation}
\begin{equation}
T(z,t) =  T_0+ \frac{e_s}{e_i}(T_0-T_s)\cdot \erf \left(\frac{z}{2\sqrt{D_{i}t}}\right)  \,\,{\rm for}\,\, 0 \leq z \leq h(t) 
\label{eq:tempprofi}
\end{equation}
where $e_{s,i}=\sqrt{\lambda_{s,i} \rho_{s,i} C_{ps,i}}$ are the effusivities of the substrate and the solid, and $T_0$ the contact temperature at the solid/substrate interface (a constant in time in this self-similar framework).
The effusivity of a material is the physical quantity that witnesses both its heat capacity and its ability to diffuse it. $T_0$ is an integration constant to be determined by the boundary conditions. It corresponds to the temperature at the solid/substrate interface and the self-similar behaviour indicates that this temperature is a constant in time! The error function, \erf\ is defined here by:
\begin{equation}
    \erf (x) = \frac{2}{\sqrt \pi} \int_0^x e^{-\xi^2} d\xi.
    \label{eq:erf}
\end{equation}
Then, by imposing the Stefan condition at $z = h(t)$ (eqs. \ref{BC:zh}), we obtain the following transcendental equation:
\begin{equation}
    \text{St}= \frac{\sqrt{\pi \beta}}{2} e^{\frac{\beta}{4}} \left(\frac{e_i}{e_s} + \erf\left(\frac{\sqrt{\beta}}{2}\right)\right)
    \label{eq:implicit}
\end{equation}
that links St, the Stefan number, with the ratio of the diffusion coefficients $\beta$, defined respectively by:
\begin{equation}
    \text{St} = \frac{C_{pi} (T_m - T_s)}{L} \,\,\,{\rm and} \,\,\, \beta=\frac{D_{\text{eff}}  }{ D_{i}} 
    \label{eq:ste}
\end{equation}
From the relation (\ref{eq:implicit}) it is easy to deduce the asymptotic behaviours for small and large Stefan numbers:
\begin{equation}
 \beta \sim \frac{4 e_s^2}{\pi e_i^2} \text{St}^2 \,\,\,\,{\rm for}\,\, \text{St}\ll1, \hskip 0.5 cm \text{and} \hskip 0.5 cm \beta \sim 4 \ln(\text{St}) \,\,\,\,{\rm for}\,\,\,\, \text{St}\gg 1.
 \label{asymp:ste}
\end{equation}
An interesting physical quantity to compute is the  dimensionless solid/substrate interface temperature $\bar T_0$ yielding:
\begin{equation}
\overline{T}_0=\frac{T_0-T_s}{T_m-T_s}=\frac{1}{1+\frac{e_s}{e_i}\erf \left(\frac{\sqrt{\beta}}{2} \right)}
    \label{eq:T0}
\end{equation}
such that $\overline{T}_0$ varies between $0$ and $1$ with $\overline{T}_0=0$ for $T_0=T_s$ and $\overline{T}_0=1$ for $T_0=T_m$ (see Fig.~\ref{ModelGeom}). Then from the asymptotic relations between $\beta$ and $\text{St}$ (Eqs.~\ref{asymp:ste}), we obtain that $T_0 \rightarrow T_m$ for St $\ll1$, 
while for St $\gg1$ it converges to an intermediate temperature:
\begin{equation}
\lim_{\text{St} \rightarrow \infty} \,T_{0} = T_s+(T_m-T_s)\frac{1}{1+\frac{e_s}{e_i}}
\label{eq:T0lSt}
\end{equation}
 which corresponds to the interface temperature when two infinite media 
 are suddenly put in contact \citep{Ruiter2018}.\\

\begin{figure}
\includegraphics[width=\hsize]{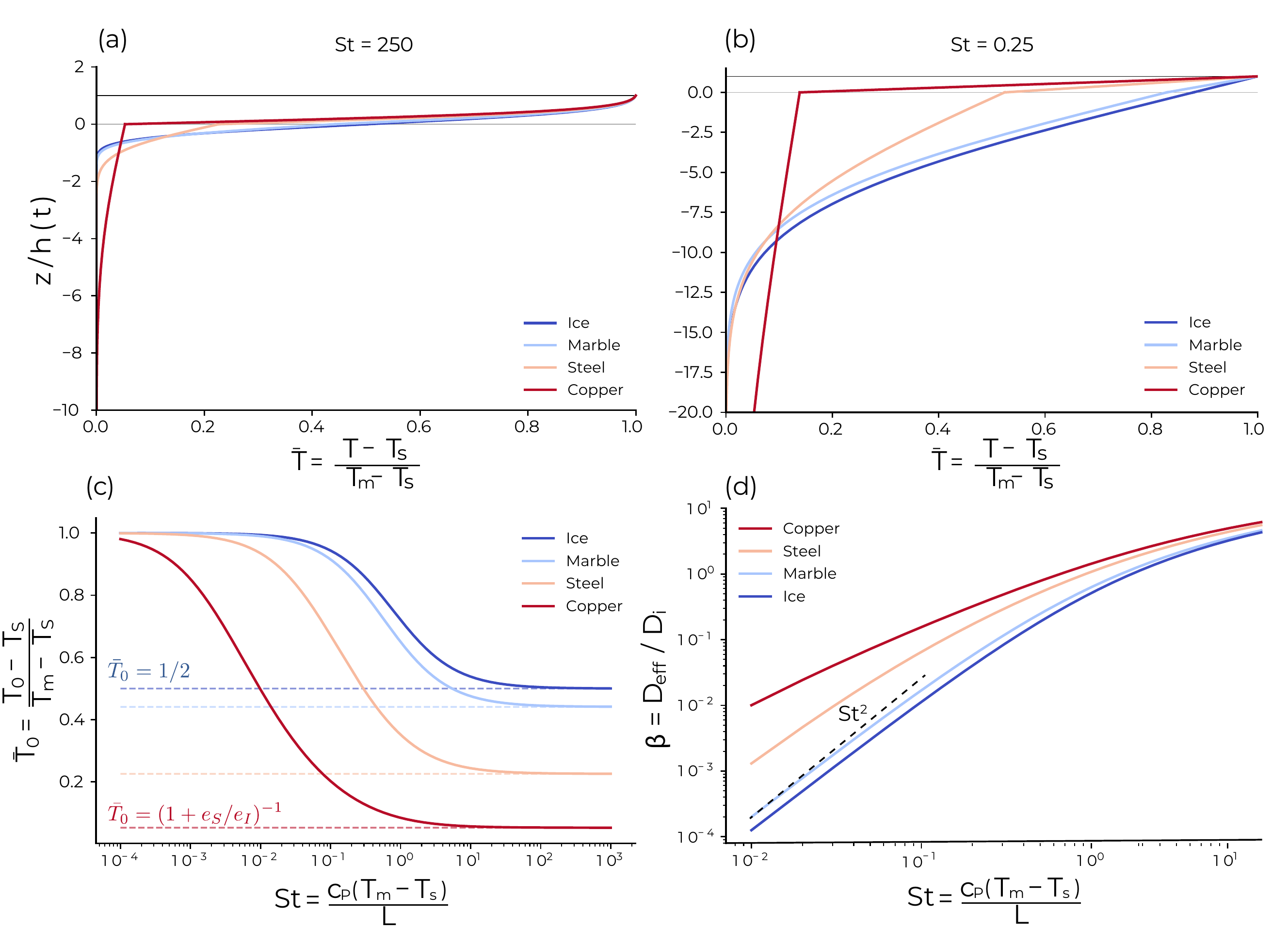}
    \caption{Results of the model for the different substrates (copper: red, steel: orange, marble: light blue) and 
    ice (dark blue) that is for comparison with the Stefan problem with infinite ice and infinite liquid water.
        (a) \& (b): Dimensionless temperature profiles $\overline{T}(z,t)= \frac{ T(z,t)-T_s}{T_m-T_s}$ as function of the self-similar variable $z/\sqrt{D_\text{eff}t}$, obtained from Eq. \ref{eq:tempprofi} and \ref{eq:tempprofs} for two characteristic Stefan number $\text{St}=250$ (a) and $\text{St}=0.25$ (b). 
        (c) Dimensionless contact temperature, obtained from Eq. \ref{eq:T0}. Regardless of the substrate, it goes from $1$ for $\text{St} \rightarrow 0$ to the asymptotical values $(1 + e_s/e_i)^{-1}$ (indicated on the figure) when $\text{St} \rightarrow \infty$, reminiscent of the  contact temperature of two infinite bodies initially at different temperatures.
        (d) The effective diffusion coefficient of the solidification front normalized by the thermal diffusion coefficient of the ice, $\beta=D_{\text{eff}}/D_{i}$, as a function of the Stefan number. The differences between substrates are more important at low Stefan number, and the asymptotic regime at low Stefan number $\beta \propto St^2$ is indicated (dashed line). 
    }
  \label{fig:model}
\end{figure}

In order to understand the thermal fields in the substrate and the solid layer, we plot the vertical position normalized by the thickness of the growing ice layer ($z/\sqrt{D_\text{eff}t}$) as a function of the normalized temperature profiles:
$$\overline{T}(z,t)= \frac{ T(z,t)-T_s}{T_m-T_s}$$
for two specific values of the Stefan number : a large value $ \text{St}=250\gg1$ (figure \ref{fig:model}(a)) and a value of order one $ \text{St}=0.25\sim1$ (figure \ref{fig:model}(b)) ; and for four couples melt/substrat. The melt is water as it is the most common in our experiments and four substrates have been chosen : copper, steel and marble in order to have a large range of thermal conductivity, and ice in order to compare our result with the \textit{classical Stefan problem} with no substrate. 
We observe that the temperature profiles evolve from the substrate temperature $T_s$ ($\overline{T}(z\!\!\rightarrow \!\!-\infty,t)=0$) up to the solidification temperature $T_m$ ($\overline{T}(h(t),t)=1$). 
%
In the high Stefan numbers regime (Fig.~\ref{fig:model} (a)), the temperature profiles exhibit the shape of the error function within both the solid and the substrate domains. The temperature gradient is discontinuous at the contact between the ice and the substrate (z=0), because of the discontinuity of the thermal conductivities. For $\text{St}=0.25$ (Fig.~\ref{fig:model} (b)), the temperature profile tends to be linear in the ice while it still exhibits an error function profile in the substrate.

Figures~\ref{fig:model} (c) and (d) present respectively the normalized temperature of the solid/substrate interface, $\overline{T}_0$ (Eq.~\ref{eq:T0}), and the normalized effective diffusion coefficient $\beta=D_\text{eff}/D_i$ (Eq.~\ref{eq:hsim}), both plotted as functions of the Stefan number, for the same four couples melt/substrate as before. 
Remember that increasing the Stefan number corresponds to decrease the substrate temperature. 


To better understand the different regimes observed in figure~\ref{fig:model}, we can investigate the two asymptotic behaviours of our model: the solidification-dominated regime (St~$\ll$~1) and the solid-cooling dominated (St~$\gg$~1).
Let us first consider the latter case of large Stefan numbers (St~$\gg$~1) that corresponds to $C_{pi}\Delta T\gg L$, meaning that the latent heat $L$ released by the solidification at the solid/liquid interface $z=h(t)$, is negligible compared to the heat energy released by cooling the solid ($C_{pi}\Delta T$) from the melting temperature T$_m$ to the substrate one T$_s$. 
In this case, 
the substrate is almost not warmed up by the solidification and does not influence the front propagation.
As a consequence, the dynamics is dominated by the self-similar variation of the temperature profiles with time; they exhibit two $\erf$ function that join at the solid/substrate interface (Fig.~\ref{fig:model} (a)). If the diffusive coefficients in the solid and the substrate are the same, which is here the case when the substrate is ice and almost the case when the substrate is marble, the $\erf$ shape of the temperature profile are symmetrical with respect to $z=0$. In other cases (steel and copper), this symmetry of the error function is broken.
In this limit, the temperature of the solid/substrate interface is the asymptotic limit given by the equation (\ref{eq:T0lSt}) (Fig.~\ref{fig:model}~(c)). 
Finally, figure~\ref{fig:model} (d) confirms that in this limit ($St\gg 1$), the front propagation dynamics is not much influenced by the substrate thermal conductivity as all the substrates gather on the \textit{classical Stefan problem} (dark blue curve) with no substrate. 


In the opposite limit ($\text{St}\ll1$), the latent heat released by the solidification is much larger than the heat released by cooling the solid down to $T_s$, indicating that the heat reaching the substrate comes mainly from the solid/liquid interface. The solid layer is inert in the sense that it only transfers the heat flux coming from the solidification to the substrate: the heat flux stays constant through the solid layer.
This explains why the temperature field is linear in the solid (Fig.~\ref{fig:model} (c)). The slope is selected by the front dynamics and evolves slowly with time, it can be estimated using Eq. (\ref{asymp:ste}) for St $\ll1$ ($\beta \sim \text{St}^2$) as follows~:
$$ \lambda_i \frac{\partial T}{\partial z}= \rho_i L \frac{dh}{dt}=\frac12 \rho_i L \sqrt{\frac{D_\text{eff}}{t}}\sim \text{St} \rho_i L \sqrt{\frac{D_i}{t}}. $$
This linear profile in the solid joins an $\erf$ profile in the substrate which goes deeper (relatively to h(t)) compared to the large Stefan regime. 
The freezing process is efficient enough to warm the substrate up. 
This explains why the interface temperature $T_0$ increases (Fig.~\ref{fig:model}(d)) and tends toward $T_m$ ($T_0 \rightarrow T_m$ for $St \rightarrow 0$ (Eq. \ref{eq:T0})). 
Finally, the effective diffusion coefficient is smaller than in the large Stefan regime (Fig.~\ref{fig:model} (d)). This can be understood easily since decreasing the Stefan number corresponds to warming the substrate up, so the ice grows slower. In this regime we observe that all the curves follow $D_\text{eff} \sim \text{St}^2 D_i$ as shown on Eq. (\ref{asymp:ste}).
What is very interesting is that $D_\text{eff}$ with metal substrates differs by several orders of magnitude from the one with no substrate (dark blue curve). Indeed, as the Stefan number decreases, the solid gets more and more inert, and the solidification front propagation becomes more and more influenced by the substrate thermal parameters. In this case, the higher the substrate thermal conductivity, the faster the solidification front propagation (see Eq. (\ref{asymp:ste})). This property demonstrates how important it is to take into account the heat transfert within the substrate.

\subsection{Experimental comparison}
\label{ExperimentalComparison}

\begin{figure}
    \includegraphics[width=\hsize]{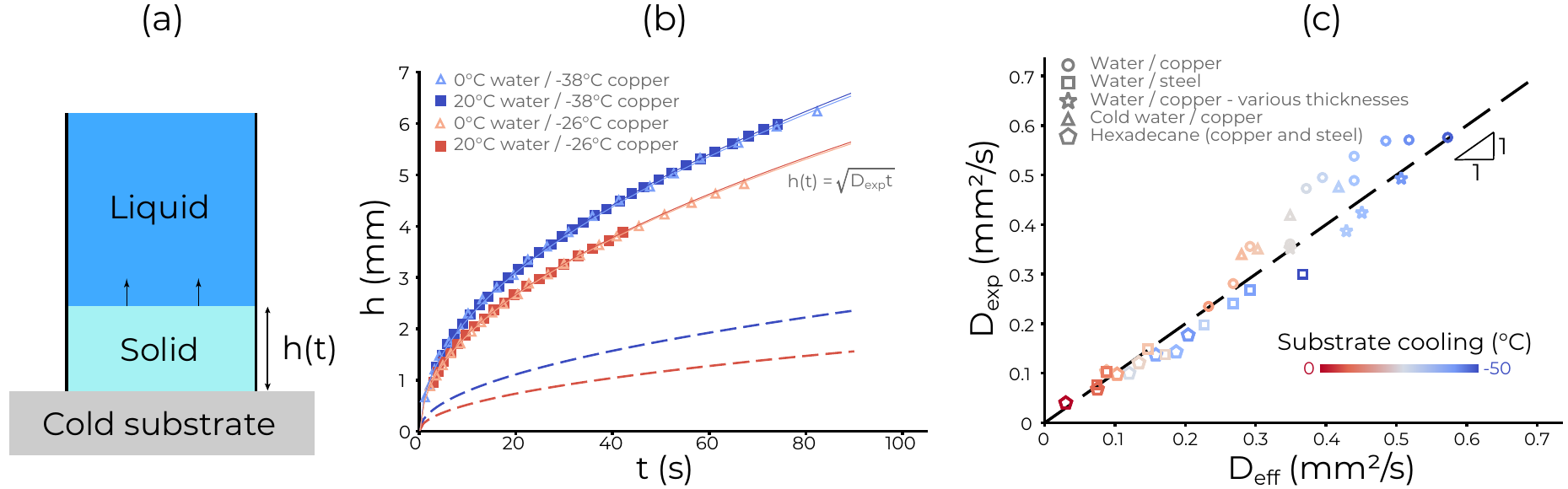}
    \caption{(a) Schematic of the experimental setup: A liquid-filled cylinder is set upon a cold substrate. The liquid starts freezing into a solid layer of thickness $h(t)$, which is monitored using a camera. 
        (b) Growth of the solid phase as a function of time. $h(t)$ is plotted in the case of water freezing on a copper plate at two different temperatures (blue and red squares), the liquid being initially at room temperature. The same experiments are repeated with water slightly warmer than 0$^\circ$C (triangles), showing no differences. The dynamic of each experiment is fitted by a square root function (full lines), in order to get its diffusion coefficient $D_\text{exp}$. Dashed lines represents the dynamic predicted by the \textit{classical Stefan problem} (without the substrate), for the two different temperatures.
(c) Comparison of the experimental freezing dynamic and the theory: The diffusion coefficients $D_\text{exp}$ and \deff, respectively experimental and theoretical, are compared for a wide range of parameters. The colour refers to the substrate cooling 
$\Delta T = T_s - T_m$. Circles (respectively squares) represents water at room temperature freezing on copper (respectively steel). Stars refer to thin thicknesses of water freezing on copper, triangles to 0$^\circ$C water freezing on copper. Finally, pentagons are for hexadecane freezing on both metals. The $y=x$ guideline is shown as a dashed line.}
  \label{TubeExp}
\end{figure}

In the aim of validating our model and its assumptions, we carry out a model experiment which consists in putting in contact a liquid with a substrate at a temperature below the liquid solidification temperature (see Fig.~\ref{TubeExp} (a)). The liquid is initially poured in a transparent PMMA tube of 3.4 cm diameter. We use two different liquids (water and hexadecane) and two different substrates (copper and steel). We also varied the initial temperature of the water ($20$ and $0^\circ$C) and the thickness of the initial water layer (between 4 millimeters and a few centimeters). Finally, the propagation of the solidification front (h(t)) is recorded using a camera, between one and ten frames per second depending on the experiment.

Figure ~\ref{TubeExp} (b) presents the growth of the ice layer (h(t)) with respect to time for water solidifying on a copper substrate. The blue and red squares are for water at 20$^\circ$C and two different temperatures of the substrate, respectively -38$^\circ$C and -26$^\circ$C as indicated on the graph. As expected the freezing front propagates faster when the substrate is colder. On the same graph, the empty triangles show the front propagation with water initially at 0$^\circ$C. Their variations coincide with the preceding ones on the whole range of time, indicating that solidifying water at 20$^\circ$C or at 0$^\circ$C gives rise to the same solidification dynamics. This important observation justifies \textit{a posteriori} the approximation to consider the liquid layer at $T_m$ in the model and to neglect the heat flux in the liquid. 
Superimposed on these four experimental curves, the lines are the best fit using a square root function of the form :
\begin{equation}
h(t) = \sqrt{D_\text{exp}t},
 \label{Dexp}
\end{equation}
with $D_\text{exp}$ an experimental diffusion coefficient that quantifies the solidification dynamics. We observe that, as shown in the model, the solidification front follows a diffusive law.  
Note that in the experiments shown here, the same diffusive law is followed up to 7mm, suggesting that the 1D approximation seems valid at least up to a one fifth aspect ratio. 
Finally, we plot on the same graph with dashed line the variation of h(t) given by the \textit{classical Stefan problem} associated to the two substrate temperatures. In these plots, the substrate is treated as if it was ice and the thermal parameters of the substrate are consequently note taken into account. It appears that, by promoting the heat transfert, 
the substrate highly increases the front propagation velocity.  

By fitting the solidification front evolution by Eq.~(\ref{Dexp}) on each experiment we obtain $D_\text{exp}$, the experimental diffusion coefficients. On Fig.~\ref{TubeExp} (c) we compare them to the effective diffusion coefficient given by our 1D model \deff, for all of our experiments : water (0, 20$^\circ$C and different liquid thicknesses) and hexadecane, on copper and steel (see the different markers on the graph), and for various substrate temperatures between 0 and -50$^\circ$C, indicated by the colour of the markers. All the data gather along the black dashed line that has a slope of 1, indicating that for each experiment $D_\text{exp}\simeq D_\text{eff}$ : our 1D model provides therefore an excellent estimation of the dynamics of the upwards solidification when a liquid is placed on the top of a cold substrate.

\section{Solidification dynamics of an impacted drop of water}
\label{DropSolidification}

With the relevant model for the solidification dynamics of a liquid put suddenly in contact with a cold substrate, let us now consider our experiment of a water drop impacting a cold surface and see whether it can predict the final thickness for the frozen impacted drop.

\subsection{Experimental setup and qualitative description}\label{sec:exp}


The classical drop impact setup consists of a syringe pump pushing a liquid through a capillary tube from which the drop falls. As the pumping is slow enough, the size of the drop is entirely controlled by the radius of the capillary tube. 
We used two different tubes of inner diameter $1600$ $\mu$m and $250$ $\mu$m leading to two drop radii: R = 1.9 mm and 1.2 mm, yielding respective volumes 30 $\mu$L and 7 $\mu$L .
The impact velocity $U_0$ is controlled by the height of fall $H$, which in our case ranges from $15$ cm to $45$ cm, so that $U_0$ ranges from 1.7 m.s$^{-1}$ to 3 m.s$^{-1}$ (following roughly $U_0=\sqrt{2gH}$). 

Our substrates, made of blocks ($100 \times 100 \times 30$ mm) of different materials (steel, copper and marble), 
are placed into a bowl and cooled down by pouring a certain amount of liquid nitrogen. The minimal temperature reached in this work is around $-80^\circ$C. 
Due to the substrate heat capacity and to the bowl thermal isolation, it takes several hours for the system to warm up to room temperature. 
The change in the substrate temperature is thus much less than $1^\circ$C during the time of an experiment which is roughly $1$ second. 

In order to minimize frost formation, the whole system is placed inside a regulated atmosphere chamber which allows us to drastically reduce the humidity (less than 1\% humidity inside the chamber). 
The substrate temperature $T_s$ is measured before each experiment using a surface thermometer. 
The dynamic of the impact is studied using a high-speed camera, and the height profile of the frozen drop is extracted with a polychromatic confocal sensor (CCS OptimaPlus from STIL Optics)  moving along a translation platform.

\begin{figure}
    \includegraphics[width=\hsize]{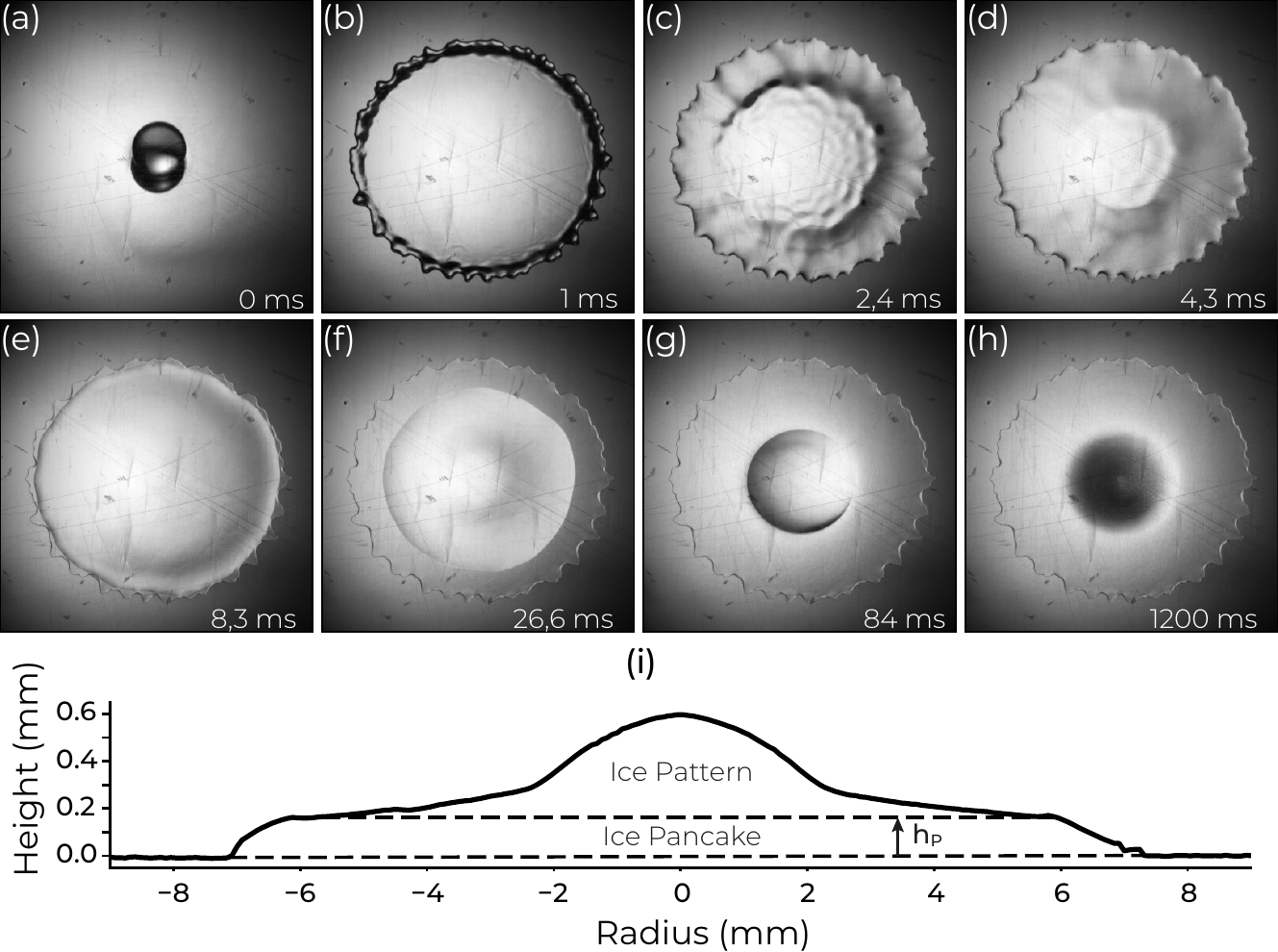}
    \caption{Snapshots (a)-(h) show a sequence at different times (indicated on each image) of a drop of water of radius 1.9 mm impacting at velocity 2.6m.s$^{-1}$ on an aluminium substrate at temperature -9$^\circ$C.
            (i) is a height profile of the final frozen splat, measured using the optical profilometry. The aspect ratio is 20, so the real splat is
             much flatter.}
  \label{fig:img}
\end{figure}

At room temperature, as a water drop impacts a solid substrate, it spreads, reaches a maximal radius \citep{Laan2014} and immediately starts to retract back to an equilibrium radius~\citep{Bartolo2005, Josserand2016}. 
Figure \ref{fig:img} (a)-(h) shows a sequence of snapshots of a water drop impacting a sub-zero substrate. Here again, the drop spreads rapidly and reaches a maximal spreading diameter (a)-(b). But then the drop does not retract, it is stuck at its maximal diameter (c). This pinning is due to the formation of a thin layer of ice between the substrate and the liquid during the spreading. On images (c)-(e) the drop is thus made of a thin layer of ice, attached to the substrate and growing vertically, beneath a thicker layer of water where capillary waves are damped. This remaining water layer is not stable due to its high aspect ratio, so that it still needs to retract in order to reach its equilibrium, and we observe on (e) where the liquid layer actually just started to depin from the edge. Then, the liquid retracts on the ice layer (f), until it reaches its typical equilibrium contact angle on ice \citep{Knight1971} and forms a spherical-cap-shaped drop (g). Eventually, it completely freezes, yielding a pointy ice drop \citep{Anderson1996, Snoeijer2012, Marin2014}, on top of an ice pancake (h). 

Figure \ref{fig:img} (i) shows the height profile of the frozen drop (h) obtained by scanning the drop diameter with our optical profilometer. This profile shows clearly two different parts: a quasi-cylindrical plate (the so-called ice pancake), whose thickness is denominated as $h_p$, on top of which we find an ice pattern, consequence of the water retraction on ice. We thus define  clearly $h_p$ as the ice layer thickness of the
edge of the structure when it depins. 
This structure with an ice pancake in contact with the impacted solid is found anytime a drop impacts a sub-zero cold surface. The ice pattern that is eventually formed on the top of the ice pancake can exhibit different shapes that will be the subject of future works. Here we focus on the ice pancake that is, in fact, crucial in all relevant applications, in particular for the stability of the frozen drop. Indeed, as the temperature of the ice varies, thermal expansion/contraction generates elastic stress that can lead to the delamination of the pancake~\citep{Ruiter2018} or the formation of cracks~\citep{Ghabache2016b}, influenced by the thickness $h_p$. The aim of the present paper is thus dedicated to the quantification of this thickness $h_p$


%
%


\subsection{The stationary contact line (SCL) regime} 
\label{sec:res}

\begin{figure}
    \includegraphics[width=\hsize]{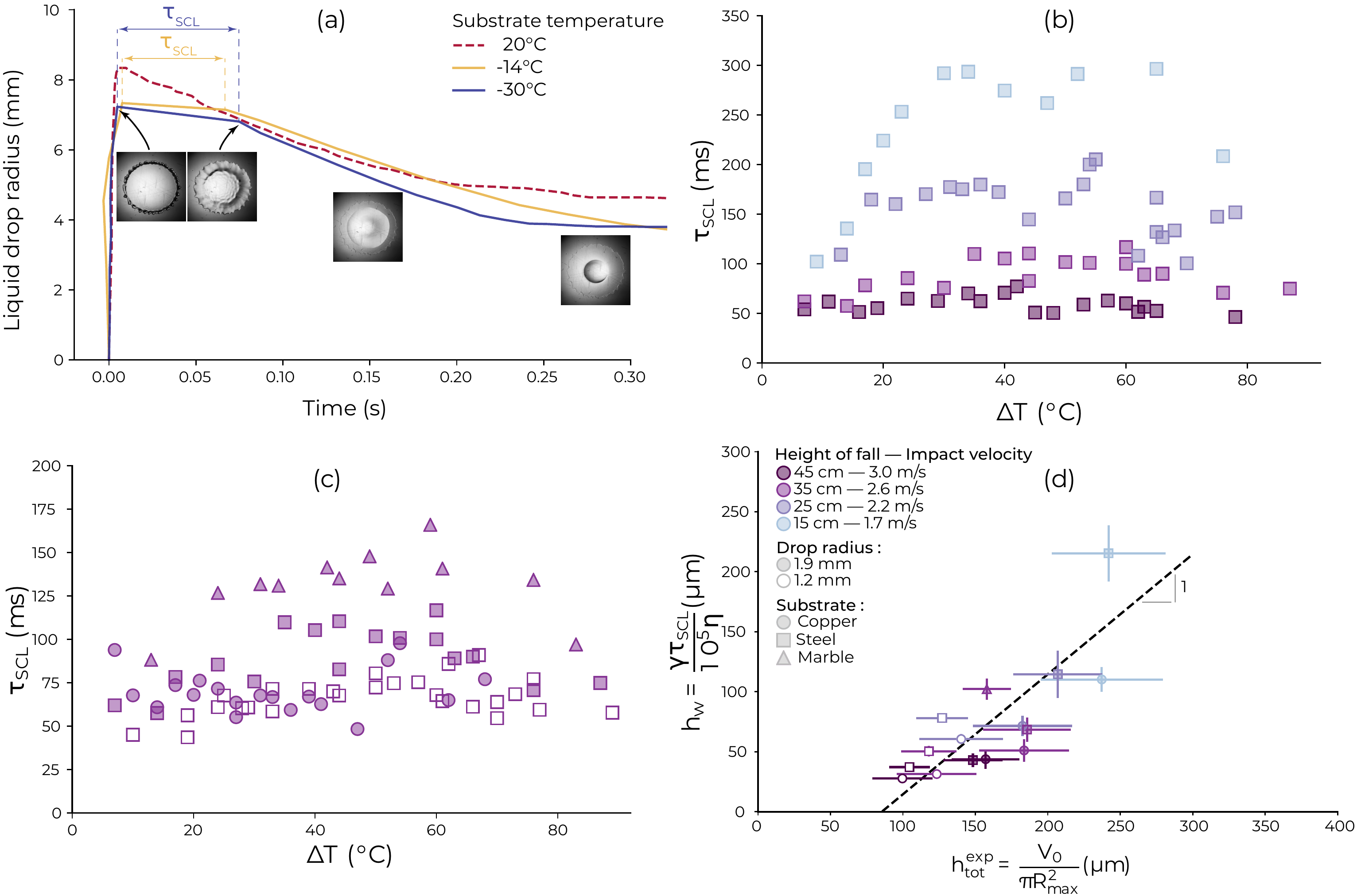}
    \caption{(a) The dynamic of spreading and retraction of the drop. The radius of the liquid film is plotted as function of time.
                Inset pictures show the aspect of the film on the different stages.
    		On all graphs, impact velocities are represented by colours (light blue for the slowest to dark purple for the fastest),
    		substrates by symbols (circles for copper, squares for steel and triangles for marble) and drop radii by the filling (full symbols for
    		$R=1.9$mm, empty symbols for $R=1.2$mm).
        (b)$\tau_{\text{SCL}}$ as a function of the temperature $\Delta T$ for different impact velocities ). The drop radius is $1.9$mm and the 			substrate is steel.
        (c)$\tau_{\text{SCL}}$as a function of the temperature $\Delta T$ for different substrates. The impact velocity is $2.6$ m/s and the 		   		drop radius is $1.9$mm (full symbols).  Results for a drop of $R=1.2$mm  impacting on steel at the same velocity are shown with 						empty squares.
        (d) Theoretical prediction for the water film thickness $h_w$ as a function of its experimental estimation $h_\text{tot}^\text{exp}$. Each point 					represents a series experiments with given impact velocity, substrate and drop radius. The error bars contain the variations within 					each series. The dashed line is a linear fit with the slope set to one.
 		}
  \label{fig:residence}
\end{figure}

As we observe on figure~\ref{fig:img}, after spreading, the contact line seems steady (b-e), before a dewetting 
transition occurs leading to the retraction of the water film on the ice layer. It is thus expected that the thickness of the ice 
pancake $h_p$ is built by thermal conduction during this time. Consequently, we need first to characterize the 
duration of this regime, that we call stationary contact line (SCL) regime \citep{Rivetti2015}. 
Figure~\ref{fig:residence} (a) presents the variation of the liquid film radius during an impact, plotted as a 
function of time, for three different substrate temperatures. The dashed line shows the liquid drop radius 
evolution at room temperature and the two solid lines at freezing température~: -14 and -30$^\circ$C. Without 
freezing (dashed line), the drop spreads rapidly, reaches its maximal diameter and almost instantaneously 
retracts. The behaviour is different when the drop freezes. Indeed, after having reached its maximal diameter 
(illustrated on the curve with the inserts), the stationary contact line regime is observed : the liquid remains 
attached to the ice layer close to its maximum radius. There, the liquid radius barely varies with time during 
approximately $60-70$ ms before the retraction of the liquid film starts. This time between the spreading and 
the retraction regimes defines the SCL time : $\tau_\text{SCL}$, as shown on the curves. 
We note that the two SCL times, for the two different temperatures, seem to be quite close.

Figure \ref{fig:residence}(b) shows the dependence of the time $\tau_\text{SCL}$ with the temperature, for 
different heights of fall and same drop size (1.9 mm) and substrate material (steel). Instead of the substrate 
temperature T$_s$,  we use $\Delta T=T_m-T_s$, where $T_m$ is the freezing temperature of the liquid (here 
for water $T_m=0^\circ$ C so that $\Delta T=-T_s$ with the Celsius scale), which is the relevant temperature 
here, because we saw that the energy needed to cool the drop down to $T_m$ can be neglected. 
The first observation is that $\tau_\text{SCL}$ always reaches a plateau where it is independent of $\Delta T$. 
The value of this plateau strongly depends on the drop impact velocity and thus on the initial spreading of the 
liquid film : the larger the maximal diameter at impact, the shorter the liquid layer stays stationary atop the ice 
pancake before retracting. These results suggest therefore that $\tau_\text{SCL}$ might be independent of the 
heat transfert and due to the sole liquid dynamics. 

Figure \ref{fig:residence}(c) presents the variation of $\tau_\text{SCL}$ with $\Delta T$ for our three different 
substrates:  marble (triangle markers), steel (square), copper (circle), and two different drop radii on steel 
substrate (full and empty squares).
In each case, the plateau regime is quickly reached confirming this independence of SCL time with $\Delta T$.  
As expected, the drop size, as the drop impact velocity, seems to play a role in the selection of the plateau 
value of $\tau_\text{SCL}$. 
What is more surprising here is the variation of the plateau value with the substrate material : $\tau_\text{SCL}$ 
decreases as the substrate thermal conductivity increases. 
This appears as a contradiction with the previous result that $\tau_\text{SCL}$ does not depend on the 
substrate temperature and, thus, seems to be independent of the thermal parameters! We will come back on 
these observations in the following.%

To explain the existence of this delay $\tau_\text{SCL}$ between spreading and retraction when water freezes, 
we will use an argument of a previous work \citep{Rivetti2015} on the relaxation of a contact line pinned at the 
edge of a polymer film. Similarly, we assume here that the time $\tau_\text{SCL}$ is due to the relaxation 
dynamics of the contact angle $\theta$ formed by the liquid film and the growing ice pancake, and pinned at its 
edge. The contact line relaxation starts at an angle given by the spreading dynamics (influenced by the drop 
impact velocity, the substrate wetting angle, and the drop size), and ends when $\theta$ reaches a critical value 
$\theta^*$ enabling its depinning, required prelude to its receding motion. In the experiment of \cite{Rivetti2015}, this angle is $\theta^* = 4.5^\circ \pm0.5^\circ$ and surprisingly appears to be independent of the liquid and of the film thickness. 

Because the liquid film is very thin, its dynamics can be taken in the lubrication regime so that the contact line relaxation is expected to follow a capillaro-viscous relaxation time $t_{w} \propto (h_0 \eta)/\gamma$, with the proportional coefficient being a function of $\theta^*$, being about $10^5$ in their experiments. Note that, as soon as the contact line retracts, they showed that $\theta$ increases to a receding contact angle which stays roughly constant during dewetting.

In order to show that the same relaxation dynamics is at play in our experiments, we compare on figure \ref{fig:residence} (d) the characteristic thickness $h_w$ of the liquid film that corresponds to the relaxation time $\tau_\text{SCL}$ following this capillaro-viscous dynamics, {\it i.e.} $h_w= \gamma\tau_\text{SCL}/(10^5\eta)$, and the characteristic thickness of the water film $h_\text{tot}^\text{exp}$ formed at the end of the spreading, neglecting ice formation  : $h_\text{tot}^\text{exp} = V_\text{tot}/\pi R_\text{max}^2$ with $V_\text{tot}$ the total volume of the drop and $R_\text{max}$ the spreading radius. Remarkably, the experimental data gather along a line of slope $1$, suggesting that this relaxation scenario of the contact line is correct. Note that, since $h_\text{tot}^\text{exp}$ corresponds to the total height of the spread drop, it is the sum of the ice pancake and the water film, so that the intersection of the dotted line with the x-axis gives a consistent estimation of the ice pancake thickness. We point out the fact that the thicknesses of the ice and liquid layer are variying in time, so that $h_\text{tot}^\text{exp}$ is just a rough approximation of the liquid film thickness. This might explain the data scattering around the linear prediction.

We have used for comparison the same value of $\theta^*$ than for the polymer films (leading to the $10^5$ prefactor for $h_w$), while there is no reason that this value is relevant for water films. In fact, when trying to fit the best line for the data of figure~\ref{fig:residence}(d), we found an angle slightly different but with no significant improvement when comparing with the $\theta^*$ correlation, so that we have kept this value for the sake of simplicity. The fact that this dewetting angle $\theta^*$ appears to be almost independent on the substrates, the liquids and the film thicknesses, while our experiments and those of ~\cite{Rivetti2015} concern film with totally different thickness 
is a very interesting and intriguing result. It might shed light on an universal dewetting mechanism that should deserve specific investigations in the future.

Finally, following these results, the stationary contact line time ($\tau_\text{SCL}$) varies linearly with the thickness of the liquid pancake at impact ($h_\text{tot}^\text{exp}$). As the spread drop diameter increases with the impact velocity, the liquid pancake is thinner and we understand the decrease of $\tau_\text{SCL}$ when the impact velocity increases, observed on figure~\ref{fig:residence} (b). 
These results resolve also the contradiction that we draw concerning the dependance of $\tau_\text{SCL}$ with the different substrate. 
The contact line relaxation dynamics suggests indeed that the substrate intervenes through its wetting properties that influences the spreading radius, and therefore the liquid film thickness pinned on the thin ice layer. In this scenario, the thermal properties of the substrate plays no direct role, explaining thus the independence of $\tau_\text{SCL}$ with the temperature.

\subsection{The ice pancake}\label{sec:results}

\begin{figure}
    \includegraphics[width=\hsize]{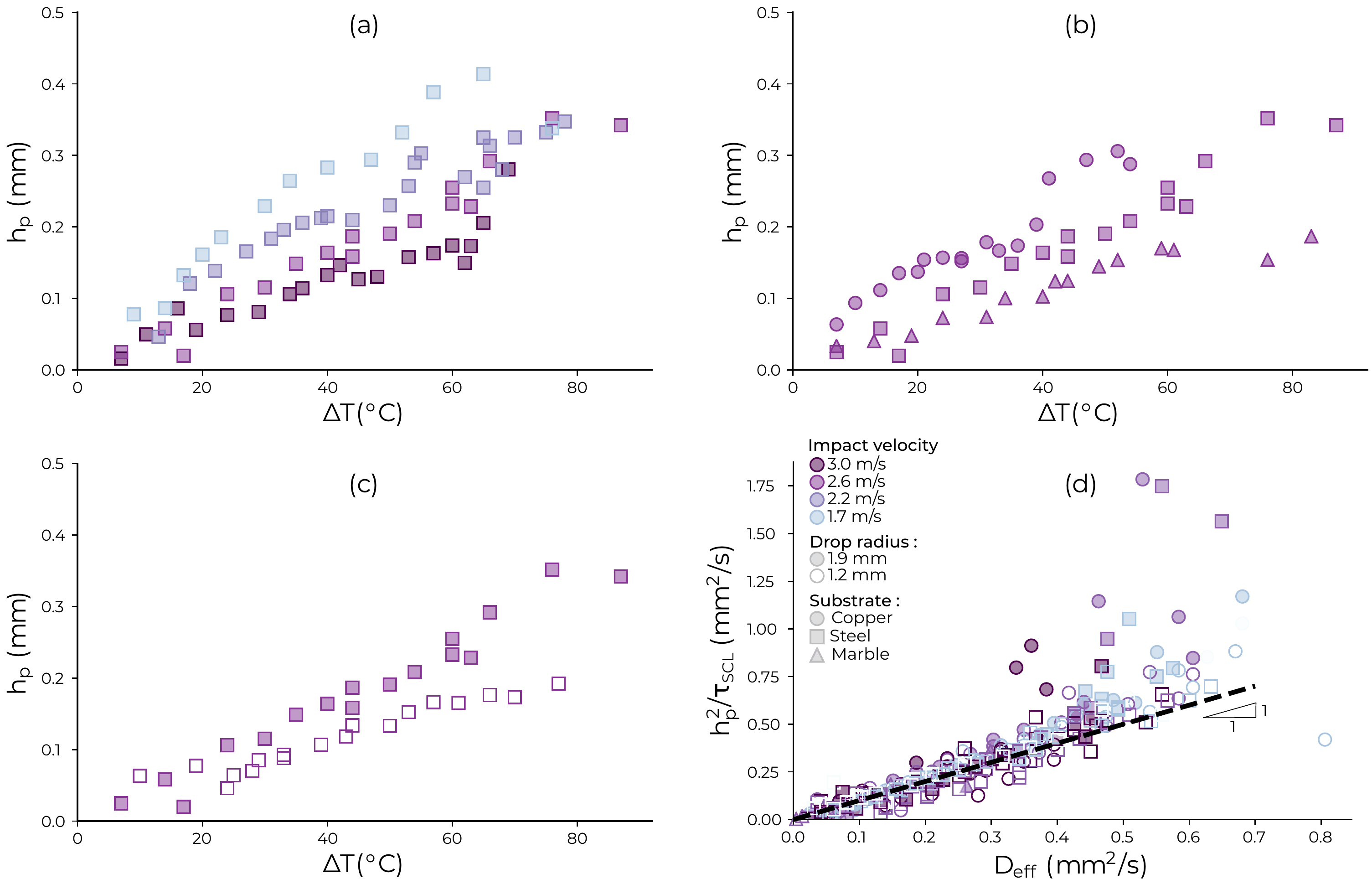}
    \caption{        
        (a) Evolution of the underlying ice plate thickness $h_p$ as a function of the temperature $\Delta T$ for different impact velocities, on the same substrate (steel)
        and with the same drop radius (1.9mm). The colder the substrate is, the thicker the ice layer is. 
        For a given temperature the ice layer is thinner for the highest impact velocities.
        (b) Evolution of $h_p$ along $\Delta T$,
        for different substrates, with the same impact velocity (2.6 m/s) and drop radius (1.9 mm).
        For a given temperature, the more conductive the substrate, the thicker the ice layer.
        (c) Evolution of $h_p$ with $\Delta T$ for different drop radius, on the same substrate (steel) and at the same velocity 
        (2.6 m/s). The smaller the drop the less it freezes.
        (d) Rescaling of the measurements against the model: the measured diffusivity of the solidification front, $h_p^2 / \tau_{SCL}$
        is plotted against the effective diffusion coefficient obtained through the previously described model (Eq. \ref{eq:hsim} and 
        \ref{eq:implicit}). 
       The dashed line has a slope 1.
        }
  \label{fig:final}
\end{figure}

At this point, we know the time during which the ice pancake is building ($\tau_\text{SCL}$) and we have a validated model for the ice growth dynamics ($D_\text{eff}$); we now need the experimental measurements of the pancake thickness: $h_p$. 
Figure~\ref{fig:final} (a)-(c) show the ice thickness $h_p$, deduced from the height profiles of the frozen drops (see Fig. \ref{fig:img} (i)), as a function of $\Delta T$, respectively for four different drop impact velocities (a), three different substrates (b) and  two different drop sizes (c). These control parameters, and their corresponding markers, are the same as those used in figure~\ref{fig:residence}. 

These graphs show that the underlying ice layer becomes thicker when:
(i) the substrate is colder (Fig.~\ref{fig:final} (a), (b) and (c)); which is expected since more liquid can be frozen during the solidification time that appeared to be mostly constant with $\Delta T$ (Fig.~\ref{fig:residence} (b) and (c)). 
(ii) The substrate heat conductivity increases (Fig.~\ref{fig:final} (b)): the heat is indeed transferred to the substrate with higher efficiency, so that the freezing front can propagate faster during a solidification time that, again, does not vary much (Fig.~\ref{fig:residence} (d)). 
(iii) The drop impact velocity is slower (Fig.~\ref{fig:final}(a)), or the drop size is smaller (Fig.~\ref{fig:final}(c)), 
because the spread drop stays freezing longer before retracting (Fig.~\ref{fig:residence} (b) and (c)).

We can now compare our experimental results to our theoretical model. Figure \ref{fig:final} (d) presents the variation of $h_p^2/\tau_\text{SCL}$, which represents the experimental diffusion coefficient of the solidification front, as a function of the theoretical diffusion coefficient $D_\text{eff}$, for all the substrate materials, substrate temperatures, drop sizes and impact velocities investigated experimentally. The substrate temperature ranges from -5$^\circ$C to -80$^\circ$C, which yields the Stefan number ranging from approximately 10$^{-2}$ to 1. We observe a nice collapse of all the data into a straight line of slope 1. 
We can therefore conclude that the ice pancake thickness is well described by the expression :
$$h_p^2 \simeq D_{\text{eff}} \tau_{\text{SCL}}$$
This indicates clearly that our simplified 1D model accounts well for this problem and that the ice pancake formation is controlled by the thermal diffusion during the depinning time of the remaining liquid film.


We observe that some experiments are not collapsing on the line with the others. They appear particularly for an effective diffusive coefficient greater than 3.10$^{-7}$ m$^2$/s. This limit of our model probably comes from the hypothesis of semi-infinite water in the freezing model. Indeed, if we estimate the thickness of the water film at impact $h_\text{tot}^\text{exp}$ with the drop volume $V_\text{tot}$ and the spreading radius $R_\text{max}$, we find $h_\text{tot}^\text{exp} \sim V_\text{tot}/\pi R^2_\text{max}\sim 200\mu$m, with a drop of volume $V_\text{tot} = 30\mu$L spreading to a radius of $R_\text{max} = 7$mm. This thickness is smaller than the highest values of h$_p$, meaning that almost the whole water drop freezes during $\tau_{\text{SCL}}$ and the hypothesis of semi-infinite water does not hold anymore.

\section{Conclusion}\label{sec:conclusion}

By studying precisely the freezing of a drop impacting a cold substrate, we have shown that the dynamics 
could be broken down into four phases. 
(1) A rapid spreading of the drop that ends with a very thin ice layer on top of which a liquid water layer is pinned. 
(2) A stillness period during which the water layer is almost at rest, the ice layer growing by thermal exchange, and the contact angle of the liquid layer relaxing. 
(3) Then, after the time $\tau_\text{SCL}$, when the contact angle has reached a depinning threshold, the water film retracts on the ice layer until it reaches an equilibrium state. 
(4) The last phase consists in the freezing of the remaining liquid, forming the final ice pattern. 
The paper is devoted to the characterization of the thermal dynamics during the second phase time $\tau_\text{SCL}$ when the contact line is pinned. Our goal is to determine the subsequent thickness of the ice pancake, $h_p$.


We have firstly developed a unidimensional solidification model that considers the heat diffusion in the solid and in the ice coupled with the Stefan condition for the solidification front. A dedicated experiment has been run in order to validate this model, showing its relevance and checking its hypotheses, in particular the one-dimensional geometry, the liquid initially taken at $T_m$, and the semi-infinite liquid phase. This leads to the characterization of the diffusion front dynamics, that involves the effective diffusion coefficient D$_\text{eff}$.

We have observed that, over a large range of temperature, $\tau_\text{SCL}$ is independent of the temperature and of the substrate thermal parameters, proving that this phase is controlled by the contact angle relaxation. Therefore, we have shown that $\tau_\text{SCL}$ can be roughly expressed by the unique expression~: 
$$\tau_\text{SCL} = 10^5\frac{\eta}{\gamma}\left(70+\frac{V_\text{tot}}{\pi R_\text{max}^2}\right)$$
%
%
Using the model, we provide a general expression for the ice pancake thickness :
$$h_p^2 \simeq  D_{\text{eff}} \tau_{\text{SCL}}=10^5 D_{\text{eff}} \frac{\eta}{\gamma}\left(70+\frac{V_\text{tot}}{\pi R_\text{max}^2}\right)$$
that accounts quantitatively for almost all of our experiments. This quantity $h_p$ has many practical interests since it provides an estimate of the splat thickness formed by the impact. Controlling this thickness is thus crucial for coating and 3D printing technology~\citep{Lipson2013} and for predicting the further mechanical behaviour of the splat~\citep{Ghabache2016b,Ruiter2018}. Our work provides therefore a general framework to model and study more complex configuration such as multiple drops impacts for airplane icing or spray coating~\citep{Chandra09} for instance.

\bibliographystyle{jfm}
\bibliography{biblio,goutte}

\begin{thebibliography}{44}
\expandafter\ifx\csname natexlab\endcsname\relax\def\natexlab#1{#1}\fi
\def\au#1{#1} \def\ed#1{#1} \def\yr#1{#1}\def\at#1{#1}\def\jt#1{\textit{#1}}
  \def\bt#1{#1}\def\bvol#1{\textbf{#1}} \def\vol#1{#1} \def\pg#1{#1}
  \def\publ#1{#1}\def\arxiv#1{#1}\def\org#1{#1}\def\st#1{\textit{#1}}

\bibitem[Allmen \& Blatter(2013)]{Allmen2013}
{\sc \au{Allmen, Martin~v} \& \au{Blatter, Andreas}} \yr{2013} {\em Laser-beam
  interactions with materials: physical principles and applications\/}, ,
  \vol{vol.~2}.  \publ{Springer Science \& Business Media}.

\bibitem[Anderson {\em et~al.\/}(1996)Anderson, Worster \& Davis]{Anderson1996}
{\sc \au{Anderson, DM}, \au{Worster, M~Grae} \& \au{Davis, SH}} \yr{1996}
  \at{The case for a dynamic contact angle in containerless solidification}.
  \jt{Journal of crystal growth}  \bvol{163}~(3),  \pg{329--338}.

\bibitem[Bartolo {\em et~al.\/}(2005)Bartolo, Josserand \& Bonn]{Bartolo2005}
{\sc \au{Bartolo, Denis}, \au{Josserand, Christophe} \& \au{Bonn, Daniel}}
  \yr{2005}  \at{Retraction dynamics of aqueous drops upon impact on
  non-wetting surfaces}.  \jt{J. Fluid Mech.}  \bvol{545},  \pg{329--338}.

\bibitem[Baumert {\em et~al.\/}(2018)Baumert, Bansmer, Trontin \&
  Villedieu]{Baumert2018}
{\sc \au{Baumert, A}, \au{Bansmer, S}, \au{Trontin, P} \& \au{Villedieu, P}}
  \yr{2018}  \at{Experimental and numerical investigations on aircraft icing at
  mixed phase conditions}.  \jt{International Journal of Heat and Mass
  Transfer}  \bvol{123},  \pg{957--978}.

\bibitem[Brillouin(1930)]{Brillouin1930}
{\sc \au{Brillouin, Marcel}} \yr{1930} Sur quelques probl{\`e}mes non
  r{\'e}solus de la physique math{\'e}matique classique propagation de la
  fusion.  \bt{In {\em Annales de l'institut Henri Poincar{\'e}\/}}, ,
  \vol{vol.~1},  \pg{pp. 285--308}.

\bibitem[Cao {\em et~al.\/}(2009)Cao, Jones, Sikka, Wu \& Gao]{Cao2009}
{\sc \au{Cao, Liangliang}, \au{Jones, Andrew~K}, \au{Sikka, Vinod~K}, \au{Wu,
  Jianzhong} \& \au{Gao, Di}} \yr{2009}  \at{Anti-icing superhydrophobic
  coatings}.  \jt{Langmuir}  \bvol{25}~(21),  \pg{12444--12448}.

\bibitem[Chandra \& Fauchais(2009)]{Chandra09}
{\sc \au{Chandra, S.} \& \au{Fauchais, P.}} \yr{2009}  \at{Formation of solid
  splats during thermal spray deposition}.  \jt{J. Thermal Spray Tech.}
  \bvol{18},  \pg{148--180}.

\bibitem[Cline \& Anthony(1977)]{Cline1977}
{\sc \au{Cline, HE} \& \au{Anthony, TRf}} \yr{1977}  \at{Heat treating and
  melting material with a scanning laser or electron beam}.  \jt{Journal of
  Applied Physics}  \bvol{48}~(9),  \pg{3895--3900}.

\bibitem[De~Ruiter {\em et~al.\/}(2017)De~Ruiter, Colinet, Brunet, Snoeijer \&
  Gelderblom]{De-Ruiter2017}
{\sc \au{De~Ruiter, Rielle}, \au{Colinet, Pierre}, \au{Brunet, Philippe},
  \au{Snoeijer, Jacco~H} \& \au{Gelderblom, Hanneke}} \yr{2017}  \at{Contact
  line arrest in solidifying spreading drops}.  \jt{Physical Review Fluids}
  \bvol{2}~(4),  \pg{043602}.

\bibitem[Dhiman \& Chandra(2005)]{Chandra05}
{\sc \au{Dhiman, R.} \& \au{Chandra, S.}} \yr{2005}  \at{Freezing-induced
  splashing during impact of molten metal droplets with high weber numbers}.
  \jt{Int. J. Heat Mass Transf.}  \bvol{48},  \pg{5625--5638}.

\bibitem[Dhiman {\em et~al.\/}(2007)Dhiman, McDonald \& Chandra]{Dhiman2007}
{\sc \au{Dhiman, Rajeev}, \au{McDonald, Andr{\'e}~G} \& \au{Chandra, Sanjeev}}
  \yr{2007}  \at{Predicting splat morphology in a thermal spray process}.
  \jt{Surface and Coatings Technology}  \bvol{201}~(18),  \pg{7789--7801}.

\bibitem[Fauchais {\em et~al.\/}(2004)Fauchais, Vardelle, Vardelle \&
  Fukumoto]{Fauchais2004}
{\sc \au{Fauchais, P}, \au{Vardelle, A}, \au{Vardelle, M} \& \au{Fukumoto, M}}
  \yr{2004}  \at{Knowledge concerning splat formation: an invited review}.
  \jt{Journal of Thermal Spray Technology}  \bvol{13}~(3),  \pg{337--360}.

\bibitem[Font {\em et~al.\/}(2017)Font, Afkhami \& Kondic]{Font2017}
{\sc \au{Font, Francesc}, \au{Afkhami, Shahriar} \& \au{Kondic, Lou}} \yr{2017}
   \at{Substrate melting during laser heating of nanoscale metal films}.
  \jt{International Journal of Heat and Mass Transfer}  \bvol{113},
  \pg{237--245}.

\bibitem[Gao \& Sonin(1994)]{Gao1994}
{\sc \au{Gao, Fuquan} \& \au{Sonin, Ain~A}} \yr{1994}  \at{Precise deposition
  of molten microdrops: the physics of digital microfabrication}.  \jt{Proc. R.
  Soc. Lond. A}  \bvol{444}~(1922),  \pg{533--554}.

\bibitem[Ghabache {\em et~al.\/}(2016)Ghabache, Josserand \&
  S{\'e}on]{Ghabache2016b}
{\sc \au{Ghabache, Elisabeth}, \au{Josserand, Christophe} \& \au{S{\'e}on,
  Thomas}} \yr{2016}  \at{Frozen impacted drop: From fragmentation to
  hierarchical crack patterns}.  \jt{Physical Review Letters}  \bvol{117}~(7),
  \pg{074501}.

\bibitem[Griffiths(2000)]{Griffiths2000}
{\sc \au{Griffiths, RW}} \yr{2000}  \at{The dynamics of lava flows}.
  \jt{Annual Review of Fluid Mechanics}  \bvol{32}~(1),  \pg{477--518}.

\bibitem[Gupta(2003)]{Gupta03}
{\sc \au{Gupta, S.C.}} \yr{2003} {\em The Classical Stefan Problem - Basic
  Concepts, Modelling and Analysis\/}.  \publ{Elsevier Science B.V, Amsterdam}.

\bibitem[Hauk {\em et~al.\/}(2015)Hauk, Bonaccurso, Roisman \&
  Tropea]{Roisman15}
{\sc \au{Hauk, T.}, \au{Bonaccurso, E.}, \au{Roisman, I.V.} \& \au{Tropea, C.}}
  \yr{2015}  \at{Ice crystal impact onto a dry solid wall. particle
  fragmentation}.  \jt{Proc. R. Soc. A}  \bvol{471},  \pg{20150399}.

\bibitem[Huppert(1986)]{Huppert1986}
{\sc \au{Huppert, Herbert~E}} \yr{1986}  \at{The intrusion of fluid mechanics
  into geology}.  \jt{Journal of Fluid Mechanics}  \bvol{173},  \pg{557--594}.

\bibitem[Huppert(1989)]{huppert1989}
{\sc \au{Huppert, Herbert~E}} \yr{1989}  \at{Phase changes following the
  initiation of a hot turbulent flow over a cold solid surface}.  \jt{Journal
  of Fluid Mechanics}  \bvol{198},  \pg{293--319}.

\bibitem[Jones(1996)]{Jones1996}
{\sc \au{Jones, Kathleen~F}} \yr{1996}  \bt{Ice accretion in freezing rain.}
  {\em Tech. Rep.\/}.  \org{COLD REGIONS RESEARCH AND ENGINEERING LAB HANOVER
  NH}.

\bibitem[Jones(1998)]{Jones1998}
{\sc \au{Jones, Kathleen~F}} \yr{1998}  \at{A simple model for freezing rain
  ice loads}.  \jt{Atmospheric research}  \bvol{46}~(1-2),  \pg{87--97}.

\bibitem[Josserand \& Thoroddsen(2016)]{Josserand2016}
{\sc \au{Josserand, C} \& \au{Thoroddsen, ST}} \yr{2016}  \at{Drop impact on a
  solid surface}.  \jt{Annual Review of Fluid Mechanics}  \bvol{48},
  \pg{365--391}.

\bibitem[Knight(1971)]{Knight1971}
{\sc \au{Knight, Charles~A}} \yr{1971}  \at{Experiments on the contact angle of
  water on ice}.  \jt{Philosophical magazine}  \bvol{23}~(181),  \pg{153--165}.

\bibitem[Kreder {\em et~al.\/}(2016)Kreder, Alvarenga, Kim \&
  Aizenberg]{Kreder2016}
{\sc \au{Kreder, Michael~J}, \au{Alvarenga, Jack}, \au{Kim, Philseok} \&
  \au{Aizenberg, Joanna}} \yr{2016}  \at{Design of anti-icing surfaces: smooth,
  textured or slippery?}  \jt{Nature Reviews Materials}  \bvol{1}~(1),
  \pg{15003}.

\bibitem[Laan {\em et~al.\/}(2014)Laan, de~Bruin, Bartolo, Josserand \&
  Bonn]{Laan2014}
{\sc \au{Laan, Nick}, \au{de~Bruin, Karla~G}, \au{Bartolo, Denis},
  \au{Josserand, Christophe} \& \au{Bonn, Daniel}} \yr{2014}  \at{Maximum
  diameter of impacting liquid droplets}.  \jt{Physical Review Applied}
  \bvol{2}~(4),  \pg{044018}.

\bibitem[Lam{\'e} \& Clapeyron(1831)]{Lame1831}
{\sc \au{Lam{\'e}, G} \& \au{Clapeyron, BP}} \yr{1831} M{\'e}moire sur la
  solidification par refroidissement d'un globe liquide.  \bt{In {\em Annales
  Chimie Physique\/}}, ,  \vol{vol.~47},  \pg{pp. 250--256}.

\bibitem[Langer(1980)]{Langer1980}
{\sc \au{Langer, James~S}} \yr{1980}  \at{Instabilities and pattern formation
  in crystal growth}.  \jt{Reviews of Modern Physics}  \bvol{52}~(1),  \pg{1}.

\bibitem[Lipson \& Kurman(2013)]{Lipson2013}
{\sc \au{Lipson, Hod} \& \au{Kurman, Melba}} \yr{2013} {\em Fabricated: The new
  world of 3D printing\/}.  \publ{John Wiley \& Sons}.

\bibitem[Marin {\em et~al.\/}(2014)Marin, Enriquez, Brunet, Colinet \&
  Snoeijer]{Marin2014}
{\sc \au{Marin, Alvaro~G}, \au{Enriquez, Oscar~R}, \au{Brunet, Philippe},
  \au{Colinet, Pierre} \& \au{Snoeijer, Jacco~H}} \yr{2014}  \at{Universality
  of tip singularity formation in freezing water drops}.  \jt{Physical review
  letters}  \bvol{113}~(5),  \pg{054301}.

\bibitem[Neufeld {\em et~al.\/}(2010)Neufeld, Goldstein \&
  Worster]{Neufeld2010}
{\sc \au{Neufeld, Jerome~A}, \au{Goldstein, Raymond~E} \& \au{Worster, M~Grae}}
  \yr{2010}  \at{On the mechanisms of icicle evolution}.  \jt{Journal of Fluid
  Mechanics}  \bvol{647},  \pg{287--308}.

\bibitem[Nishinaga(2014)]{Nishinaga2014}
{\sc \au{Nishinaga, Tatau}} \yr{2014} {\em Handbook of crystal growth:
  fundamentals\/}.  \publ{Elsevier}.

\bibitem[Pasandideh-Fard {\em et~al.\/}(2002)Pasandideh-Fard, Pershin, Chandra
  \& Mostaghimi]{Pasandideh-Fard2002a}
{\sc \au{Pasandideh-Fard, M}, \au{Pershin, V}, \au{Chandra, S} \&
  \au{Mostaghimi, J}} \yr{2002}  \at{Splat shapes in a thermal spray coating
  process: simulations and experiments}.  \jt{Journal of Thermal Spray
  Technology}  \bvol{11}~(2),  \pg{206--217}.

\bibitem[Rivetti {\em et~al.\/}(2015)Rivetti, Salez, Benzaquen, Rapha{\"e}l \&
  B{\"a}umchen]{Rivetti2015}
{\sc \au{Rivetti, Marco}, \au{Salez, Thomas}, \au{Benzaquen, Michael},
  \au{Rapha{\"e}l, Elie} \& \au{B{\"a}umchen, Oliver}} \yr{2015}  \at{Universal
  contact-line dynamics at the nanoscale}.  \jt{Soft Matter}  \bvol{11}~(48),
  \pg{9247--9253}.

\bibitem[Rubinstein(1971)]{Rubinstein1971}
{\sc \au{Rubinstein, L.I.}} \yr{1971} {\em The Stefan Problem\/},
  \st{Translations of Mathematical Monographs},  \vol{vol.~27}.  \publ{American
  Mathematical Soc.}

\bibitem[de~Ruiter {\em et~al.\/}(2018)de~Ruiter, Soto \& Varanasi]{Ruiter2018}
{\sc \au{de~Ruiter, Jolet}, \au{Soto, Dan} \& \au{Varanasi, Kripa~K}} \yr{2018}
   \at{Self-peeling of impacting droplets}.  \jt{Nature Physics}
  \bvol{14}~(1),  \pg{35}.

\bibitem[Schiaffino \& Sonin(1997)]{Schiaffino1997a}
{\sc \au{Schiaffino, Stefano} \& \au{Sonin, Ain~A}} \yr{1997}  \at{Motion and
  arrest of a molten contact line on a cold surface: an experimental study}.
  \jt{Physics of Fluids (1994-present)}  \bvol{9}~(8),  \pg{2217--2226}.

\bibitem[Schremb {\em et~al.\/}(2018)Schremb, Roisman \& Tropea]{Schremb2018}
{\sc \au{Schremb, Markus}, \au{Roisman, Ilia~V} \& \au{Tropea, Cameron}}
  \yr{2018}  \at{Normal impact of supercooled water drops onto a smooth ice
  surface: experiments and modelling}.  \jt{Journal of Fluid Mechanics}
  \bvol{835},  \pg{1087--1107}.

\bibitem[Snoeijer \& Brunet(2012)]{Snoeijer2012}
{\sc \au{Snoeijer, Jacco~H} \& \au{Brunet, Philippe}} \yr{2012}  \at{Pointy
  ice-drops: How water freezes into a singular shape}.  \jt{American Journal of
  Physics}  \bvol{80}~(9),  \pg{764--771}.

\bibitem[Stefan(1891)]{Stefan}
{\sc \au{Stefan, J.}} \yr{1891}  \at{{\"U}ber die theorie der eisbildung,
  insbesondere {\"u}ber die eisbildung im polarmeere}.  \jt{Ann. Physik Chemie}
   \bvol{42},  \pg{269--286}.

\bibitem[Tavakoli {\em et~al.\/}(2014)Tavakoli, Davis \&
  Kavehpour]{Tavakoli2014}
{\sc \au{Tavakoli, Faryar}, \au{Davis, Stephen~H} \& \au{Kavehpour, H~Pirouz}}
  \yr{2014}  \at{Spreading and arrest of a molten liquid on cold substrates}.
  \jt{Langmuir}  \bvol{30}~(34),  \pg{10151--10155}.

\bibitem[Vidaurre \& Hallett(2009)]{Vidaurre09}
{\sc \au{Vidaurre, G.} \& \au{Hallett, J.}} \yr{2009}  \at{Particle impact and
  breakup in aircraft measurement}.  \jt{J. Atmos. Ocean. Technol.}  \bvol{26},
   \pg{972--983}.

\bibitem[Viskanta(1988)]{Viskanta1988}
{\sc \au{Viskanta, R}} \yr{1988}  \at{Heat transfer during melting and
  solidification of metals}.  \jt{Journal of Heat Transfer}  \bvol{110}~(4b),
  \pg{1205--1219}.

\bibitem[Worster(2000)]{Worster2000}
{\sc \au{Worster, Michael~Grae}} \yr{2000}  \at{Solidifcation of fluids}.
  \jt{Perspectives in fluid dynamics: a collective introduction to current
  research} .

\end{thebibliography}


\end{document}